# Measurements and Characterisation of Surface Scattering at 60 GHz

Angelos A. Goulianos[1], Alberto L. Freire[1], Tom Barratt[1], Evangelos Mellios[1], Peter Cain[2], Moray Rumney[2], Andrew Nix[1] and Mark Beach[1]

[1]Communication Systems & Networks Group
University of Bristol, Merchant Venturers Building, Woodland Road, Bristol BS8 1UB, UK
Email: {A.Goulianos, M.A.Beach}@bristol.ac.uk

[2]Keysight Technology, 5 Lochside Avenue, Edinburgh Park, Edinburgh, EH12 9DJ
{peter_cain, moray_rumney}@keysight.com

*Abstract*— **This paper presents the analysis and characterization of the surface scattering process for both specular and diffused components. The study is focused on the investigation of various building materials each having a different roughness, at a central frequency of 60GHz. Very large signal strength variations in first order scattered components is observed as the user moves over very short distances. This is due to the small-scale fading caused by rough surface scatterers. Furthermore, it is shown that the diffused scattering depends on the material roughness, the angle of incidence and the distance from the surface. Finally, results indicate that reflections from rough materials may suffer from high depolarization, a phenomenon that can potentially be exploited in order to improve the performance of mm-Wave systems using polarization diversity.**

## I. INTRODUCTION

With the advancement of handheld technology, the ever growing demand for wireless connectivity continues at pace. Currently carriers are struggling to meet the service demands of their existing networks despite access to additional frequencies as a result of analogue TV switch off and the release of military bands. In addition, with LTE now having more than 31 different frequency bands globally, this is causing new issues for 4G network operators due to segmented spectrum. In order to meet the user rates demanded for 5G, new technologies are required. One leading solution for 5G and beyond networks is the use of Millimetre Wave (mm-Wave) spectrum (30-300GHz). This is now possible due to advancements in small, low cost and low power CMOS devices. This spectrum can offer large continuous bandwidths that allow multi-gigabit per second data rates. The 60GHz mm-Wave frequency band, in particular, offers a continuous block of 7GHz spanning between 57 and 64GHz in most parts of the world [1].

Currently the IEEE 802.11ad standard has created considerable interest in the use of the 60GHz unlicensed spectrum and has significantly increased the demand for 60GHz on-chip transceivers that offer up to 2GHz of bandwidth and yields theoretical speeds of up to 7Gbps, with several devices now commercially available. This marks the beginning of commercial devices that utilise the mm-Wave bands for wireless communication. Understanding, and subsequently modelling, the wireless channel is of crucial importance during the design, development and deployment process of all wireless communication systems, as it determines the quality of the link and hence the Quality-of-Service of the system as a whole. This is particularly important in the mm-Wave bands, where channel losses are significant [2]-[5] and accurate beam-forming/tracking is essential.

One of the biggest issues in the mm-Wave bands is the modelling of diffuse scattering. At lower frequencies (< 6GHz) building and terrain surfaces have usually been assumed to be electrically smooth (i.e. their surface height variations were small compared to the carrier wavelength) and the reflection process is dominated by a strong specular path at an angle of reflection equal to the angle of incidence. However, in the mm-wave bands surface height variations are significant compared to the carrier wavelength and the reflection process needs to be replaced with diffuse scatter. When a wave impinges on a rough surface scattered waves result in all directions (i.e. they are no longer restricted to the angle of reflection). The surface effectively re-radiates the incoming wave. The scattering behaviour of mm-Waves, particularly at 60GHz, has attracted significant interest from the research community. An analysis of the reflection and transmission properties of common building materials is presented in [6], while an electrical characterisation study of different materials based on power measurements is reported in [7]. Reflection measurements for interior structures of office buildings at 60GHz is presented in [8], including an investigation into the impact of circular polarisation on reflection coefficients. A scattering and reflection measurement campaign is demonstrated in [9] with a focus on extracting Lambert's Law scattering coefficients and complex permittivity coefficients for a number of different building materials.

This paper presents the results of a measurement campaign that targets the investigation of the specular and diffused scattering process for a number of different smooth and rough building materials, at 60GHz. Furthermore, the effect of depolarization is presented and analysed for each of the surfaces under test. The remainder of this paper is organized as follows: Section II provides a thorough description of the measurement campaign. Section III presents the characterization of specular reflections. Sections IV and V, provide an analysis on the diffused scattering results and the depolarization respectively. Finally, Section VI summarizes the key findings of this work and provides directions for future work.

## II. MEASUREMENT SET-UP AND EQUIPMENT

### A. Description of Specular Measurements

A 2 GHz wide baseband signal is generated using Keysight M9099 Waveform Creator software and used to configure a Keysight M8190A Arbitrary Waveform Generator (ARB) [10]. This was then employed used to drive I & Q ports of a SiversIMA up-converter, thus producing a 60 GHz modulated carrier. At the receiver, a SiversIMA device down-converts the 60 GHz signal to an IQ IF signal and this is then captured and processed using a high performance Digital Oscilloscope (DSO), MSOS804A. The Keysight 89600 VSA & Waveform Creator channel sounding function operates by repeatedly transmitting a single carrier signal bearing a modulated waveform. The waveform has excellent auto-correlation properties, and a low peak-to-average power ratio. Spectrum shaping can be applied through the use of various baseband filters available, to reduce out of band interference

For the presented measurements, the transmitting antenna was a vertically polarized rectangular horn antenna, while a circular horn was connected though an orthomode transducer at the receiver, thus allowing simultaneous horizontal and vertical polarizations to be recorded. Both antennas had a gain of 25 dBi and a 3 dB beamwidth of 13°.

In order to perform the scattering measurements, the transmitter and the receiver were mounted on a 1.00x0.70x0.72m trolley using two poles. Both the transmitter and the receiver antennas were fixed at a height of 1.5 meters and directed at the same point on the wall creating a triangle among the antennas and the surface under test. Given the large 2GHz transmission, time gating was applied to isolate the reflected specular component. It is worth mentioning that the channel sounder has been configured by an external trigger, which is sourced from a shaft encoder mounted on a wheel of the trolley. This creates a trigger every 4mm, with the equipment capable of recording at a sample rate above waking pace. Fig. 1 illustrates the hardware used during the measurements and Fig.2 depicts the measurement set-up for the characterization of specular reflections.

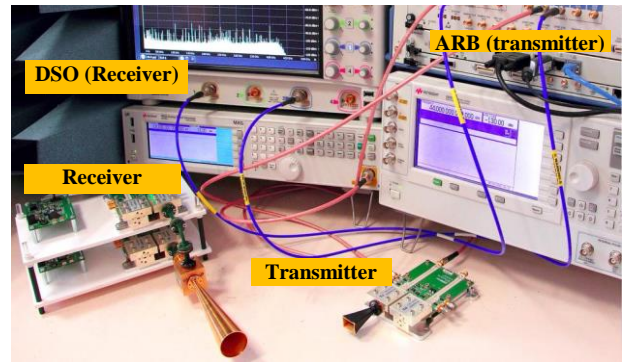

Figure 1. Measurement equipment based on Keysight channel sounder.

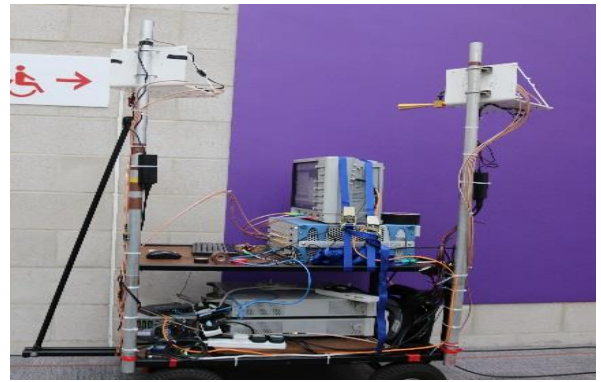

Figure 2. Measurement set-up for the characterization of specular scattering.

### B. Description of Diffused Scattering Measurements

To measure the non-specular diffuse scattered multipath components, the wideband channel sounder was pulled along in the shape of an arc around both smooth and rough building surfaces. By employing the same shaft encoder used in the specular measurements, a received impulse response was measured and recorded every 4 mm, in the range of $0^0$ to $90^0$ from the normal to the wall.

All measurements were performed at 60 GHz with a bandwidth of 2 GHz and similarly to the specular scenario, time gating was applied for the purpose of isolating the reflected component. Furthermore, the transmitting and receiving antennas were constantly pointing towards the center of the arc. Fig.3 illustrates a graphical representation of the arc measurements for a representative angle of $45^0$.

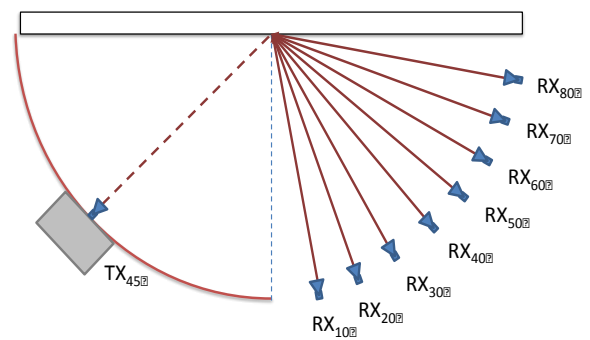

Figure 3. Graphical representation of diffused scattering arc measurements for a $45^0$ incident angle.

In order to evaluate and assess the potential dependency of diffused scattering on the transmitter-receiver separation, the angle of incidence (transmission angle) and the surface type, two building materials have been considered in this study: The red stone wall (which represents a rough wall surface) and a concrete pillar which corresponds to a smooth wall surface. The measurements carried out at an incidence angle of $15^0$, $30^0$ and $45^0$, whereas the radius of the measured arc was set to 2, 4 and 6 meters respectively.

III. CHARACTERIZATION OF SCATTERING AT SPECULAR DIRECTIONS

For the characterization of the reflected components at the specular directions, three different building materials are presented and characterized in this work. These include a very rough outdoors surface (dressed stone wall), an outdoors surface of medium roughness (bath stone wall), and a smooth surface of non-metalised window. All the measured surfaces are described and illustrated in Table I.

Table I. Measured building materials for analyzing specular scattering.

| Building Type Walls | Name | Location (Indoor/Outdoor) |
|---|---|---|
| 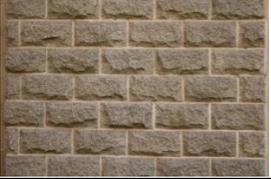 | Dressed Stone | Outdoor |
| 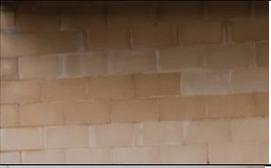 | Bath Stone | Outdoor |
| 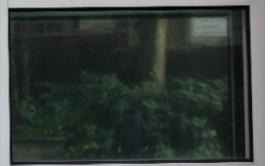 | Glass | Indoor and Outdoor |

Fig. 4 depicts the measured received power of the first order scattered component for a travel distance of 4m in parallel to the dressed stone wall (rough surface) and the glass window (smooth surface).

As observed from the Fig. 4, the impact of surface roughness on the received signal strength implies that the expected power resulting from reflections off building materials cannot be modelled only with deterministic models such as the Fresnel equations. The scattering off rough surfaces results in additional signal components that are being added constructively or destructively, creating an effect similar to the multipath small-scale fading that is observed in typical multipath-dense wireless channels. Therefore, this random fluctuation of the received signal should be superimposed on the expected deterministic specular component that is predicted by classic electromagnetic theory.

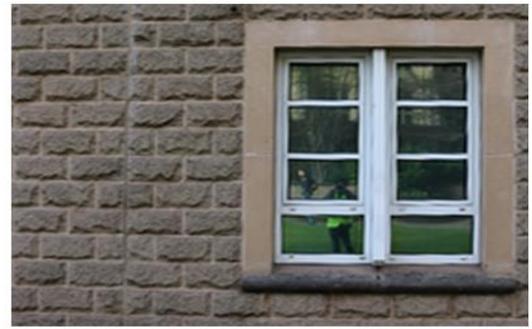
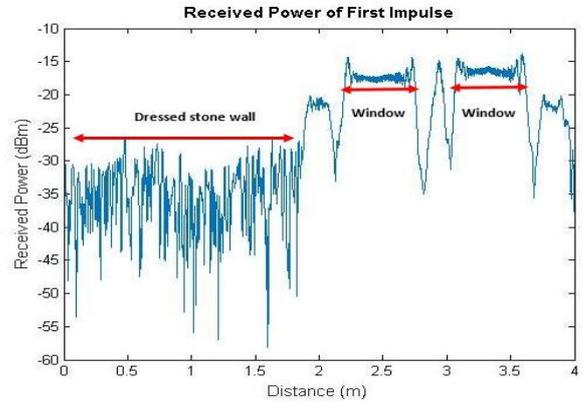

Figure 4. Measured scattering profile for the dressed stone wall and the glass window.

IV. CHARACTERIZATION OF DIFFUSED SCATTERING MEASUREMENTS

For the characterization of diffused scattering, a very smooth and a very rough surface have been considered. The red stone wall surface is illustrated in Fig. 5 and presents similar characteristics to the dressed stone wall that has been employed for the case of specular reflections. Furthermore, the smooth surface is represented by a smooth concrete pillar. The selection of both materials is related to their physical location that allows the realization of arc measurements up to a radius of 6 meters.

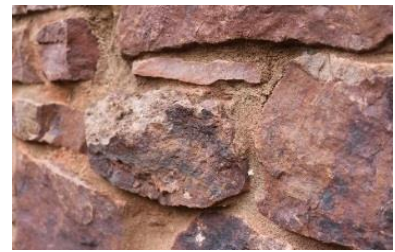

Figure 5. The red stone scattering surface.

For the purpose of determining the concentration of power within a specified range we introduce the term of power concentration which can be defined as the angular span corresponding to 90% of the power in the angular profile. Therefore, the higher the power concentration, the less the impact of the diffused scattering, since a low angular span implies that most of the received power is found within a small range around the incident angle. Fig. 6 and Fig.7 depict the power angular profiles resulting from diffused scattering measurements at a representative transmit angle of $45^0$, for the concrete pillar and the red stone wall scenarios.

The above figures illustrate the important differences observed when comparing the angular scattering profiles of rough and smooth surfaces. More specifically, it is apparent that for the case of reflections from the concrete pillar, there is a clear separation of the angle range where most of the power is concentrated. Clearly this angular span is centred at the angle of incidence. Furthermore, it has been observed that an increase in the transmitter-receiver separation decreases the angular spread for the same incident angles. In addition, it is shown that keeping the distance fixed, the power concentration is reduced for higher transmit angles.

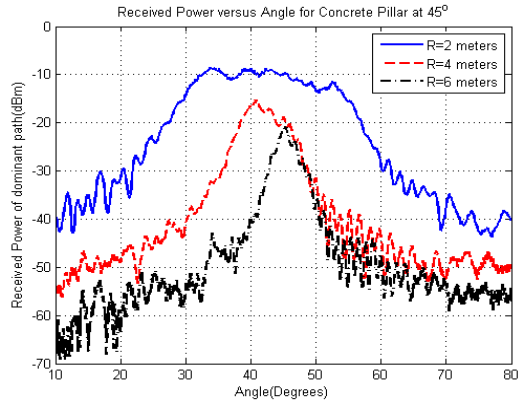

Figure 6. Power angular profiles at $45^0$ transmit angle, for the red stone wall measurements.

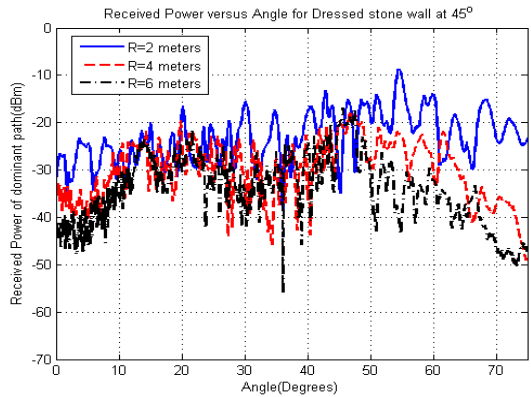

Figure 7. Power angular profiles at $45^0$ transmit angle, for the red stone wall measurements.

For the case of reflections from the red stone wall, the power angular profile appears sparse, making thus difficult to distinguish and visualize the angular span corresponding to the power concentration metric, which obviously is significantly bigger than the case of a smooth wall surface. After analysing the results, it was found that similar to the case of the concrete pillar, an increase in distance implies reduction of the angular spread. Furthermore, higher transmit angles lead to smaller values of angular spread when there is a reference at the same distance. In general, power concentration reduces with distance and transmitting angle. However, when considering reflections from rough surfaces, such as the red stone wall, the value of this metric appears quite large, implying that the effect of diffused scattering is significant and should be incorporated into electromagnetic-based prediction tools such as ray tracing algorithms.

## V. Depolarization

Reflections off a particular building material can significantly change the polarization of the incident waves in a manner which is proportional to the surface roughness of this material. In the measurements undertaken for this study, the transmitting signal was vertically polarized and the receiver was capable of receiving both the horizontal (cross-polar) and the vertical (co-polar) signal components through an orthomode transducer.

The effect of polarization has been investigated through means of the cross-polar discrimination ratio (XPD). The XPD is given by

$$XPD = 10 log_{10} \left(\frac{P_V}{P_H}\right) \qquad (1)$$

where $P_V$ and $P_H$ correspond to the received power of the vertical and the horizontal component respectively. Due to the property of the channel sounder used in these measurements both the horizontal and vertical components of the received signal have been recorded. Therefore, the estimation of the XPD ratio is straight forward and it is calculated directly from the measured data.

Some representative results on the depolarization of the vertically transmitted signal are shown in Fig. 8 and Fig. 9.

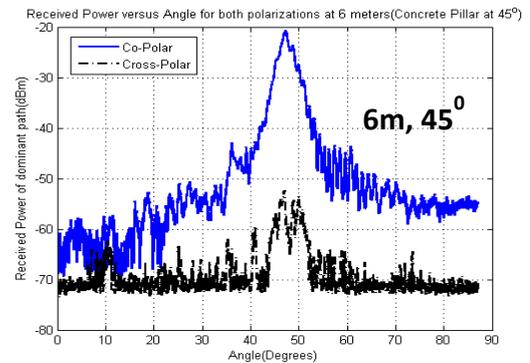

Figure 8. Scattering profiles for both vertical and horizontal polarizations at 6 meter distance and $45^0$ transmit angle – Concrete pillar.

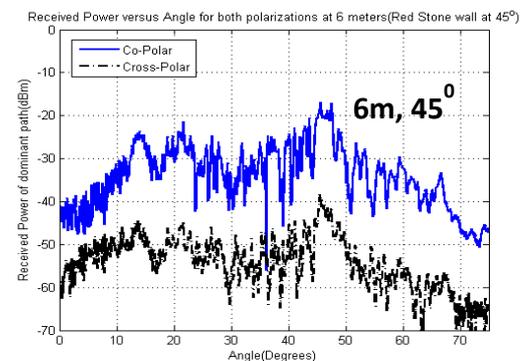

Figure 9. Scattering profiles for both vertical and horizontal polarizations at 6 meter distance and $45^0$ transmit angle – Red stone wall.

Measurement analysis suggests that as a general observation, XPD values are much lower in the case of the red stone wall. This finding implies that depolarization is more severe when considering reflections from rough wall surfaces. Another observation is that for the concrete pillar scenario, no XPD dependency on distance was observed, whereas it was shown that the cross-polar discrimination becomes bigger for higher incident angles. In contrast, the depolarization due reflections off the red stone wall has shown no dependency on distance and incident angle.

To obtain a clear illustration of the distribution of XPD ratios at the specular direction, the Empirical Cumulative Density Function (CDF) of XPD has been produced for the window, the bath stone and the dressed stone wall surface. The plot of XPD for all three materials is depicted in Fig. 10.

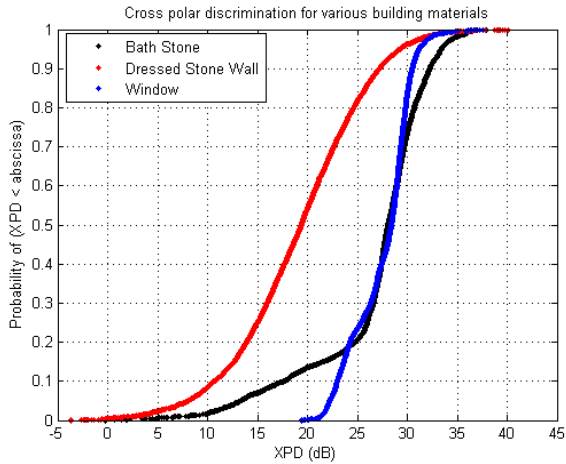

Figure 10. CDF of XPD values for window, bath stone and dressed stone wall.

Similar to the diffused scattering case, it is apparent that for the case of specular reflections, the amount of depolarization caused by a rough surface such as the dressed stone wall is significantly larger than the depolarization caused by smooth surfaces such as the window or the bath stone wall. It is worth mentioning that the XPD ratio for the dressed stone wall was at some cases smaller than 0dB. This observation implies that polarization diversity is a feature that can be adequately exploited in mm-wave frequencies, providing thus a significant improvement in the performance of mm-wave systems, especially when communication is achieved by means of dominant reflections.

## VI. CONCLUSIONS

This paper demonstrates scattering measurements and characterization at 60 GHz, for both specular and diffused components of various building materials at 60GHz. Very large signal strength variations in the first order scattered components have been observed over very short distances, for rough building surfaces. These small-scale variations could potentially have a significant effect on the system-level performance (e.g. on beam-forming and beam-tracking algorithms, on link-adaptation algorithms and on the performance and efficiency of the MAC and Network layer TCP protocols).

Furthermore, results indicate that the impact of diffused scattering becomes less severe as the distance from the scattering surface and the angle of incidence increases, when directional antennas are employed at both the transmitter and receiver side. This observation becomes more apparent for the case of relatively smooth scattering surfaces. In addition, it is shown that reflection off rough materials may suffer from high depolarization, a phenomenon that can potentially be exploited in order to improve the performance of mm-Wave systems using polarization diversity.

Results presented in this paper can form the basis for the development of an "add-on module" in order to extend current deterministic (e.g. ray-tracing), stochastic (e.g. 3GPP Spatial Channel Model) or hybrid (e.g. METIS) channel models to support spatial consistency of the diffuse scattering in expected use cases.

ACKNOWLEDGMENTS

The research leading to these results received funding from the European Commission H2020 program under grant agreement n°671650 (5G PPP mmMAGIC project).